\documentclass[fleqn,usenatbib]{mnras}

\usepackage{newtxtext,newtxmath}

\usepackage[T1]{fontenc}
\usepackage{tikz}

\DeclareRobustCommand{\VAN}[3]{#2}
\let\VANthebibliography\thebibliography
\def\thebibliography{\DeclareRobustCommand{\VAN}[3]{##3}\VANthebibliography}

\usepackage{graphicx}	
\usepackage{amsmath}	
\usepackage{xcolor}
\usepackage{caption}

\title[GALEX and carbon dredge-up in white dwarfs]{The ubiquity of carbon dredge-up in hydrogen-deficient white dwarfs as revealed by GALEX}

\author[Blouin et al.]{
Simon Blouin$^{1}$\thanks{E-mail: sblouin@uvic.ca},
Mukremin Kilic$^{2}$,
Antoine Bédard$^{3}$,
Pier-Emmanuel Tremblay$^{3}$
\\
$^{1}$Department of Physics and Astronomy, University of Victoria, Victoria, BC V8W 2Y2, Canada\\
$^{2}$Homer L. Dodge Department of Physics and Astronomy, University of Oklahoma, 440 W. Brooks St, Norman, OK 73019, USA \\
$^{3}$Department of Physics, University of Warwick, Coventry, CV4 7AL, UK\\
}

\date{Accepted for publication in MNRAS Letters on July 25, 2023}

\pubyear{2023}

\begin{document}
\label{firstpage}
\pagerange{\pageref{firstpage}--\pageref{lastpage}}
\maketitle
\begin{abstract}
The convective dredge-up of carbon from the interiors of hydrogen-deficient white dwarfs has long been invoked to explain the presence of carbon absorption features in the spectra of cool DQ stars ($T_{\rm eff} < 10{,}000\,$K). It has been hypothesized that this transport process is not limited to DQ white dwarfs and also operates, albeit less efficiently, in non-DQ hydrogen-deficient white dwarfs within the same temperature range. This non-DQ population is predominantly composed of DC white dwarfs, which exhibit featureless optical spectra. However, no direct observational evidence of ubiquitous carbon pollution in DC stars has thus far been uncovered. In this Letter, we analyze data from the \textit{Galaxy Evolution Explorer} (GALEX) to reveal the photometric signature of ultraviolet carbon lines in most DC white dwarfs in the $8500\,{\rm K} \leq T_{\rm eff} \leq 10{,}500\,$K temperature range. Our results show that the vast majority of hydrogen-deficient white dwarfs experience carbon dredge-up at some point in their evolution.
\end{abstract}

\begin{keywords}
stars: abundances -- stars: atmospheres -- white dwarfs
\end{keywords}

\maketitle

\section{Introduction}
\label{sec:intro}
Most white dwarfs are expected to possess a relatively thick hydrogen layer ($M_{\rm H}/M_{\star} \simeq 10^{-4}$, \citealt{iben1984,renedo2010}) above their helium envelopes and dense cores. This thick hydrogen layer ensures that these stars maintain a pure-hydrogen atmosphere throughout their evolution, keeping other elements well below the photosphere. However, as a result of a late helium-shell flash or binary evolution \citep{althaus2005,werner2006,reindl2014}, a significant minority of white dwarfs have a much smaller hydrogen content and exhibit helium-dominated atmospheres. These hydrogen-deficient stars represent around 30 per cent of the white dwarf population \citep{bergeron2001,tremblay2008,limoges2015,blouin2019c,ourique2019,mccleery2020,lopez2022,caron2023}. During the course of their evolution, transport processes such as diffusion and convection can lead to changes in their surface compositions, a phenomenon referred to as spectral evolution. This subject has been extensively explored in a vast and rich literature \citep[e.g.,][]{fontaine1987,macdonald1991,rolland2018,rolland2020,cunningham2020,bedard2022a,bedard2022b,bedard2023}.

Below an effective temperature ($T_{\rm eff}$) of 10{,}000\,K, hydrogen-deficient white dwarfs can appear as DQ stars, with a helium-dominated atmosphere polluted by traces of carbon that produce detectable molecular carbon absorption features in the optical. In most cases, the presence of carbon at the surface of those stars is understood as the result of a convective dredge-up process that transports carbon from the deep interior to the outer layers \citep{pelletier1986,camisassa2017,bedard2022b}. It is expected that this process affects the majority of hydrogen-deficient white dwarfs \citep{bedard2022b}.\footnote{Hydrogen-deficient white dwarfs descending from carbon-poor O(He) stars are a known exception \citep{reindl2014,bedard2022b}.} Yet, in the local sample, DQ white dwarfs represent only 15--20\% of the helium-atmosphere population cooler than 10{,}000\,K \citep{mccleery2020}. This apparent discrepancy can be resolved if most hydrogen-deficient white dwarfs are indeed polluted by carbon dredged-up from the core, but that only a minority possesses sufficient carbon in their atmospheres to show \textit{detectable} spectral carbon features. In other words, there would be a large population of ``DQ manqués'', white dwarfs with non-trivial but optically undetectable photospheric carbon abundances that appear as featureless DC stars.

This scenario is supported by a variety of observations. First, the lower limit of the carbon-to-helium abundance distribution of DQ white dwarfs (which varies with $T_{\rm eff}$) coincides with the optical visibility threshold for carbon spectral features \citep{dufour2005,bedard2022b}. Improbable coincidences aside, this hints to the existence of DQ-manqué white dwarfs just below the visibility limit. Secondly, it has recently been shown that the bifurcation between the A and B branches in the \textit{Gaia} white dwarf colour--magnitude diagram \citep{gaiahrd2018} can be explained by the existence of a large DQ-manqué population \citep{blouin2023,camisassa2023}. The additional free electrons introduced by the dredged-up carbon alter the colours of helium-rich atmospheres in a manner consistent with the observed bifurcation pattern. However, this remains indirect evidence for the occurrence of carbon dredge-up in DC white dwarfs, because traces of hydrogen can in principle have the same effect \citep{bergeron2019}. The \textit{Gaia} bifurcation can be interpreted as the signature of carbon dredge-up only after some assumptions are made regarding the photospheric hydrogen abundances of DC white dwarfs. It is possible to ground these assumptions on the abundance patterns observed in warmer DB(A) white dwarfs \citep{blouin2023}, but extrapolating to lower temperatures is still required.

While the aforementioned results make a compelling case for the occurrence of carbon dredge-up in DC white dwarfs, a more direct observational confirmation is still wanting. In this Letter, we use ultraviolet (UV) data from the \textit{Galaxy Evolution Explorer} (GALEX, \citealt{martin2005}) to reveal the photometric signature of the UV carbon lines of the large DQ-manqué population hidden among cool hydrogen-deficient white dwarfs. Our sample and observational data are described in Section~\ref{sec:sample}. We demonstrate the ubiquity of carbon pollution among the 8500--10{,}500\,K DC population in Section~\ref{sec:analysis} by comparing the GALEX data to model atmosphere calculations. Finally, we conclude in Section~\ref{sec:conclusion}.

\section{Sample selection}
\label{sec:sample}
Our objective is to analyze the UV photometry of DC white dwarfs that are predicted to be polluted by dredged-up carbon. To construct our sample of DC white dwarfs, we cross-matched all confirmed DC stars from the Montreal White Dwarf Database (MWDD, \citealt{dufour2017})\footnote{\url{https://montrealwhitedwarfdatabase.org/}} with the \textit{Gaia} catalogue \citep{gaiaedr3} and the GALEX catalogue of unique UV sources from the all-sky imaging survey \citep[GUVcat,][]{bianchi17}, which yielded a sample of 286 objects with FUV and NUV photometry. To mitigate the effects of interstellar reddening, which introduces significant scatter in the GALEX data, we discarded all objects located beyond a distance of 150\,pc. A small fraction of DC stars belongs to a subclass of massive and magnetic white dwarfs \citep[e.g.,][]{kilic2021} that most likely followed a different evolutionary pathway than the standard $\sim 0.6\,M_{\odot}$ hydrogen-deficient population. We removed those by taking cuts in the \textit{Gaia} colour--magnitude diagram to eliminate stars that do not lie on the B branch: we only kept objects that satisfy $11.75 < M_{\rm G}-4({\rm BP}-{\rm RP}) < 12.2\,$mag. This yields a sample of 68 DC white dwarfs with GALEX photometry. A linearity correction was applied to the GALEX photometry following \cite{wall2019}. All photometry was dereddened using $E(B-V)$ values from the Stilism database\footnote{\url{https://stilism.obspm.fr/}} and the extinction coefficients $A_{\lambda}/E(B-V)$ given by \cite{gentile2019} and \cite{wall2019}. Our sample is inevitably incomplete at high FUV magnitudes, but this is not a concern in the context of this work.

For comparison purposes, we also selected samples of the three other primary classes of hydrogen-deficient white dwarfs (DBs, DQs and DZs) using the same methodology. We do not select any DA white dwarfs as carbon dredge-up is not predicted to occur in standard hydrogen-rich stars. Note also that we limit our analysis to objects that are warm enough that Balmer lines should be detectable if hydrogen is present, which implies that we can confidently classify every star in our sample as being hydrogen deficient.

\section{Analysis}
\label{sec:analysis}
\subsection{The UV signature of carbon dredge-up in DC white dwarfs}
\begin{figure}
    \includegraphics[width=\columnwidth]{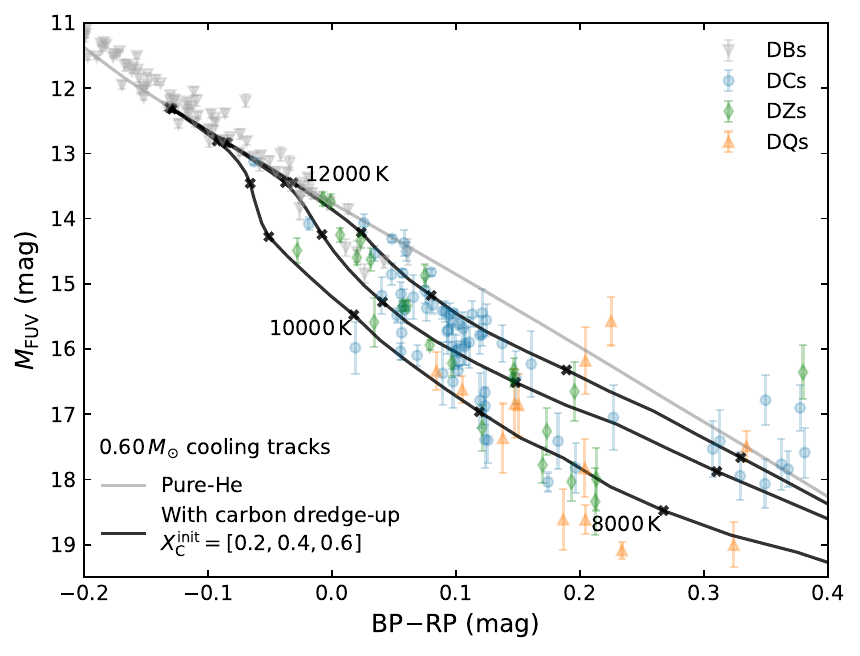}
    \caption{Colour--magnitude diagram of our sample of DB, DC, DQ and DZ white dwarfs (see legend in the upper-right corner). The grey line represents a 0.6\,$M_{\odot}$ cooling track for a pure-helium atmosphere. Black lines depict three cooling tracks for 0.6\,$M_{\odot}$ hydrogen-deficient white dwarfs that undergo carbon dredge-up, each corresponding to a different progenitor carbon abundance (the lowest sequence having the highest carbon abundance). All cooling sequences are based on theoretical calculations described in the text. Crosses along the black lines indicate increments of 1000\,K in $T_{\rm eff}$.}
    \label{fig:HR1}
\end{figure}

Figure~\ref{fig:HR1} displays our sample of hydrogen-deficient white dwarfs in a {\it Gaia}--GALEX colour--magnitude diagram. The grey line represents the theoretical cooling track for a 0.6\,$M_{\odot}$ white dwarf with a pure-helium atmosphere, based on the model atmospheres of \cite{bergeron2011,blouin2018} and the carbon--oxygen interior models of \cite{bedard2020}, assuming a thin hydrogen layer ($\log M_{\rm H}/M_{\star} = -10$). As observed in the \textit{Gaia} colour--magnitude diagram \citep{gentile2019,bergeron2019}, DB white dwarfs cluster close to the hot end of the pure-helium cooling track. A small $\sim 0.1\,$mag offset is apparent, which may be due to uncertainties inherent to the dereddening procedure and to the fact that the DB distribution possibly peaks slightly below 0.6$\,M_{\odot}$ \citep{tremblay2019b}. However, beyond ${\rm BP}-{\rm RP}=0$, where helium lines become undetectable and thus most DBs transform into DCs, the overwhelming majority of stars lie well below the pure-helium track. This deviation persists until at least ${\rm BP}-{\rm RP}=0.2$\,mag. Notably, there is a prominent concentration of objects at ${\rm BP}-{\rm RP}=0.1$\,mag that extends from 0.5 to 1.5\,mag below the pure-helium track. This is a striking discrepancy.

This behaviour is precisely what current state-of-the-art evolutionary calculations predict if we assume that most hydrogen-deficient white dwarfs undergo carbon dredge-up. The black lines in Figure~\ref{fig:HR1} show the predicted path followed by 0.6\,$M_{\odot}$ hydrogen-deficient white dwarfs once we account for carbon dredge-up. The vertical shift between the sequences with carbon dredge-up and the reference pure-helium sequence is a direct consequence of the additional opacity provided by carbon lines in the UV. These tracks were calculated as in \cite{blouin2023} by combining the element transport calculations of \cite{bedard2022b} and the model atmospheres of \cite{blouin2019c,coutu2019}.\footnote{Hydrogen traces are not included. In the 9000--10{,}000\,K temperature range, we have verified that an abundance as high as $\log\,{\rm H/He}=-5$ is insufficient to qualitatively displace the cooling tracks. $\log\,{\rm H/He}=-5$ is high enough for H$\alpha$ to be visible, which means that going above that value is irrelevant for our study of non-DA white dwarfs.} An important free parameter in these calculations is the amount of carbon in the PG~1159-like progenitors of those hydrogen-deficient white dwarfs. A range of mass fractions $X_{\rm C}^{\rm init}=0.2$ to 0.6 appears representative of existing measurements \citep{werner2006,werner2014}, and thus we show cooling tracks for $X_{\rm C}^{\rm init}$ values of 0.2, 0.4, and 0.6. We stress that the (non-constant) photospheric carbon abundances in these sequences is not fixed in some ad hoc fashion, but are directly derived from detailed numerical simulations that account for all relevant transport processes. 

Without any adjustment,\footnote{The overshoot parameter ($f_{\rm ov}=0.015$) at the bottom of the envelope convection zone has been previously calibrated to match the abundance patterns of DQ white dwarfs \citep{bedard2022b,blouin2023}.} the carbon dredge-up sequences bracket the observed distribution of DC white dwarfs strikingly well in the $0<{\rm BP}-{\rm RP}<0.15$ region. Furthermore, they also recover the concentration of DC white dwarfs positioned close to the pure-helium sequence at $0.3<{\rm BP}-{\rm RP}<0.4$. This is because most of the dredged-up carbon is predicted to have settled below the photosphere by the time that point is reached, leading to a gradual closing of the gap between the pure-helium and carbon-polluted tracks. The excellent agreement between the DC data and the dredge-up sequences (and the poor match with the pure-helium sequence) over the whole range of colours shown in Figure~\ref{fig:HR1} indicates that carbon dredge-up is a very common phenomenon for hydrogen-deficient white dwarfs.

It is noteworthy that the majority of known DQ stars are located close to the $X_{\rm C}^{\rm init}=0.6$ track. This supports the hypothesis that stars with optically detectable carbon features represent only the high-abundance tail of the carbon pollution distribution in hydrogen-deficient stars \citep{bedard2022b}.\footnote{The most flagrant outlier to this pattern, SDSS J140723.04+203918.6 ($M_{\rm FUV}=15.6$), was previously found to have an unusually low mass of 0.43\,$M_{\odot}$ \citep{coutu2019}.} Finally, we note that the DZ data cannot be interpreted without a custom model atmosphere analysis of each star, as both the externally accreted metals and the dredged-up carbon contribute to the observed FUV flux deficit (assuming the two processes are unrelated, see \citealt{blouin2022,farihi2022,hollands2022}).

While the overall agreement between the dredge-up sequences and the DC white dwarfs is remarkable, we note a possible tendency of the models to overestimate carbon abundances (and hence FUV magnitudes) between 10{,}000 and 12{,}000\,K. Indeed, almost no star lies below the $X_{\rm C}^{\rm init}=0.4$ sequences above 10{,}000\,K. We believe that this issue is another manifestation of the problem noted in \cite{blouin2023}, where dredge-up models overpredict the amplitude of the Gaia bifurcation above 10{,}000\,K. This limitation may be linked to current uncertainties in models for the average ionization level of carbon in warm dense helium, which determine how early the dredge-up process is initiated.

\subsection{Ruling out alternative explanations}
The agreement between the dredge-up predictions and the GALEX data is remarkable and represents strong observational evidence for the previously hypothesised ubiquity of carbon dredge-up in the hydrogen-deficient white dwarf population. Yet, to firmly establish this fact, we need to consider (and rule out) alternative explanations.

\begin{figure}
    \includegraphics[width=\columnwidth]{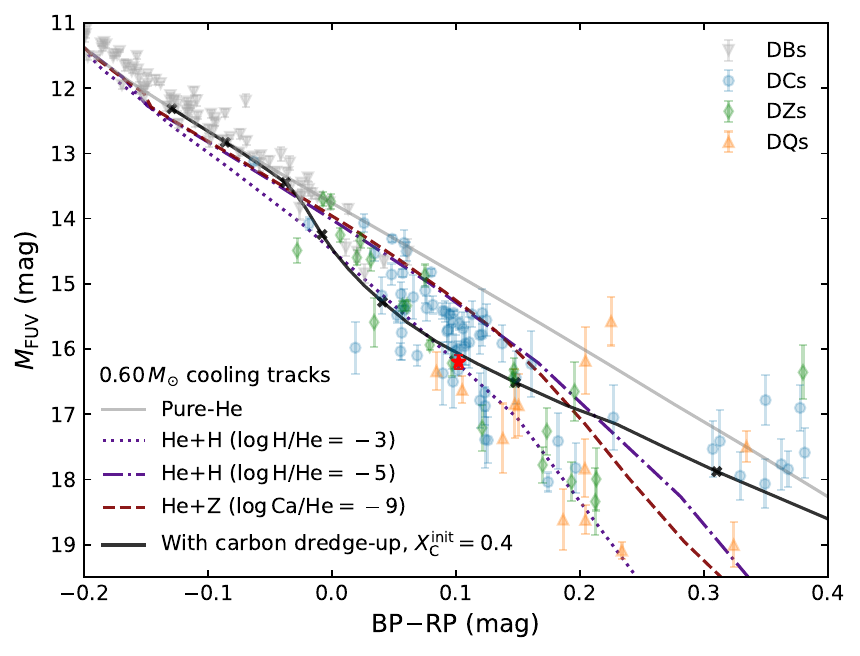}
    \caption{Identical to Figure~\ref{fig:HR1} but with different evolutionary tracks. Sequences of helium-dominated atmospheres polluted by trace amounts of hydrogen are shown in indigo for two different hydrogen abundances (see legend) and a metal-polluted cooling track is depicted in maroon. The pure-helium (grey line) and $X_{\rm C}^{\rm init}=0.4$ dredge-up (black line) sequences from Figure~\ref{fig:HR1} are included for reference. A red star highlights the object further analyzed in Figure~\ref{fig:sed}.}
    \label{fig:HR2}
\end{figure}

The accretion of planetary material onto white dwarfs is a common phenomenon \citep{koester2014}, and it can significantly affect the spectral energy distributions of hydrogen-deficient objects by introducing metals into their atmospheres \citep{dufour2007,caron2023}. Therefore, it is reasonable to consider accretion as a potential explanation for the shift between the pure-helium track and observed DC distribution in Figure~\ref{fig:HR1}. We test this hypothesis in Figure~\ref{fig:HR2}, where we show the cooling track of a helium-atmosphere white dwarf polluted by chondritic material (dashed maroon line). The photospheric calcium abundance is fixed at $\log\,{\rm Ca/He}=-9$, and the abundances of all other metals from C to Cu are assumed to have chondritic abundance ratios with respect to Ca, which is representative of most polluted white dwarfs \citep{hollands2018,blouin2020b,harrison2021,trierweiler2023}.\footnote{All abundances ratios are reported in terms of number abundances in this Letter.} While this level of pollution leads to a noticeable shift with respect to the pure-helium sequence, it is insufficient to match the observed FUV flux deficit of most DC white dwarfs. Further increasing the metal abundance is not a viable option, as at $\log\,{\rm Ca/He}=-9$ the Ca II H \& K lines are clearly visible for $T_{\rm eff} \lesssim 12{,}000\,{\rm K}$ \citep{coutu2019}, meaning that the current level of pollution is already too high to be compatible with the DC classifications of those stars. We can therefore firmly rule out accretion as a plausible explanation. Accretion of material with a very peculiar composition (e.g., extremely carbon rich) could in principle produce a sizeable FUV deficit while maintaining a featureless optical spectrum, but carbon-rich accretion is uncommon in dusty white dwarfs \citep{xu19}, and it is vanishingly improbable that \textit{all} DC white dwarfs accrete bodies with such an unusual composition.

It is well established that most hydrogen-deficient white dwarfs in this temperature range harbour traces of hydrogen in their helium-dominated atmospheres \citep[e.g.,][]{koester2015,rolland2018,genest2019,bedard2023}. Since such traces can also reduce the FUV flux due to absorption by the red wing of the Lyman-$\alpha$ line, this a priori constitutes another possible explanation for the pattern observed in Figures~\ref{fig:HR1} and~\ref{fig:HR2}. However, we can dismiss this hypothesis using a similar reasoning as in the case of accretion. In Figure~\ref{fig:HR2}, we present cooling tracks for helium-dominated atmospheres with constant hydrogen abundances of $\log\,{\rm H/He}=-5$ (indigo dash-dotted line) and $\log\,{\rm H/He}=-3$ (indigo dotted line). It is evident that abundances exceeding $\log\,{\rm H/He}>-5$ are needed to reproduce the DC distribution. However, such abundances cannot be reconciled with the DC classification of these stars, as an abundance of $\log\,{\rm H/He}=-5$ is sufficient to produce observable Balmer lines in the $T_{\rm eff} \gtrsim 9000\,$K temperature range \citep{coutu2019}. Thus, hydrogen cannot account for the observed FUV deficit, and we conclude that carbon pollution is the only possible explanation. Note that we have also tested helium-dominated models polluted by traces of hydrogen \textit{and} metals, but again found that it is impossible to match the bulk of the data while satisfying the constraint of a featureless optical spectrum. Finally, unlike carbon dredge-up, none of the alternative scenarios offer an explanation for the presence of DC white dwarfs close to the pure-helium sequence at $0.3<{\rm BP}-{\rm RP}<0.4$. 

\begin{figure}
    \includegraphics[width=\columnwidth]{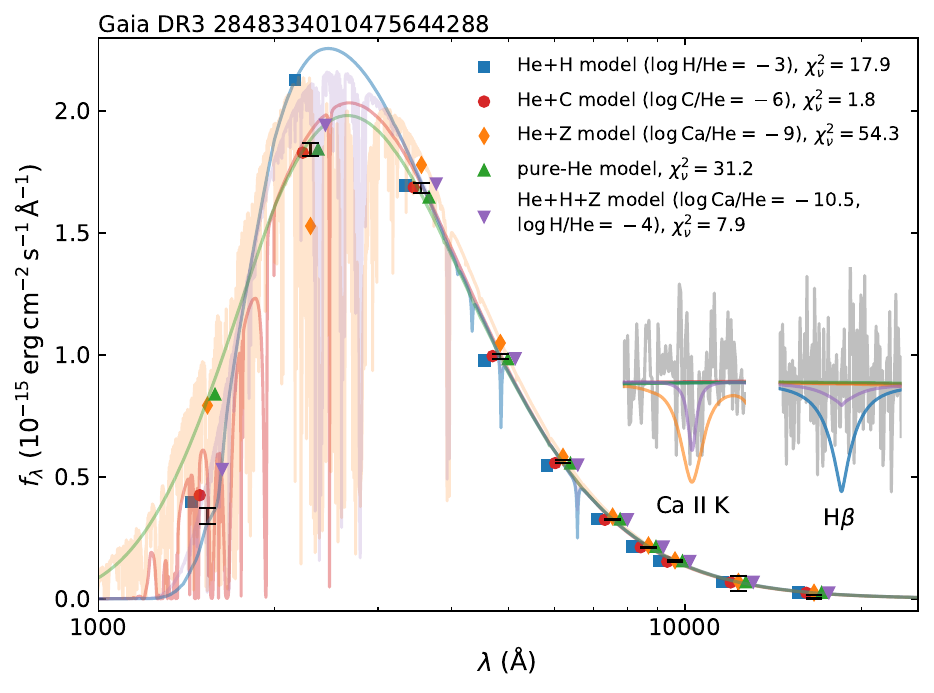}
    \caption{Observed spectral energy distribution (black error bars) of Gaia DR3 2848334010475644288, including photometry from the GALEX FUV and NUV, Sloan Digital Sky Survey (SDSS) \textit{u}, Pan-STARRS \textit{grizy} and Two Micron All Sky Survey (2MASS) \textit{JH} bands. Best-fit models assuming five different compositions are presented. The synthetic spectra are shown as solid lines and the synthetic photometry is shown as filled symbols, as described in the legend. The composition is fixed for each model, but the solid angles and effective temperatures are adjusted to minimize the $\chi^2$ between the synthetic and observed photometry (the reduced chi-squared statistics are given in the legend). For a given photometric band, the horizontal offset between symbols is for clarity only. The insets on the right compare the observed spectrum \citep{kilic2020} to the different model atmospheres in the Ca II K and H$\beta$ regions.}
    \label{fig:sed}
\end{figure}

To demonstrate more vividly the inability of metal accretion and hydrogen pollution to explain the GALEX data, we show in Figure~\ref{fig:sed} the spectral energy distribution of a typical DC white dwarf from our sample (marked in red in Figure~\ref{fig:HR2}). We overlay five model atmospheres onto the observed photometry, each having a different composition. In all five cases, the atmospheric composition is fixed, but the effective temperature and solid angle are adjusted to achieve the best fit. As anticipated, the pure-helium solution (green, $T_{\rm eff}=10{,}090\,$K) fails to reproduce the observations and flagrantly overestimates the FUV flux. Similarly, an helium-dominated atmosphere polluted by accretion (orange, $\log\,{\rm Ca/He}=-9$ and $T_{\rm eff}=10{,}090\,$K) remains too bright in the FUV band despite already having too much calcium to match the optical spectrum (see Ca II K inset).\footnote{Incidentally, the optical spectrum of Gaia DR3 2848334010475644288, classified as a DC star, shows a hint of Ca II H \& K absorption.} A model with a large trace of hydrogen (blue, $\log\,{\rm H/He}=-3$ and $T_{\rm eff}=9390\,$K) can fit the FUV band but falls short in the NUV region and produces Balmer lines that would be visible in the optical (see H$\beta$ inset). Models with both traces of hydrogen and traces of metals can fare better if the abundances are finely tuned (purple; $\log\,{\rm Ca/He}=-10.5$, $\log\,{\rm H/He}=-4$, and $T_{\rm eff}=10{,}090\,$K), but still miss the FUV and NUV bands by several standard deviations when the abundances are forced to be consistent with the optical spectroscopy. In short, only a carbon-polluted model (red) can simultaneously match all bands while appearing as a featureless DC white dwarf in the optical range. Here, the best-fit model corresponds to an effective temperature of 9420\,K, but that value is of course dependent on the uncertain carbon abundance at the photosphere.

\section{Conclusion}
\label{sec:conclusion}
We have presented strong observational evidence establishing that carbon dredge-up is a ubiquitous phenomenon for hydrogen-deficient white dwarfs. Stars with carbon features in their optical spectra represent only the tip of the iceberg: the vast majority of hydrogen-deficient white dwarfs experience the convective dredge-up of carbon at some point in their evolution. This conclusion was reached by comparing the FUV magnitudes of DC white dwarfs to model atmosphere predictions and showing that only carbon dredge-up can reproduce the observed flux depression while satisfying the constraint of a featureless optical spectrum. The excellent agreement between current dredge-up model predictions and the GALEX data further solidifies our conclusion. Our results confirm the recent claim that the \textit{Gaia} B branch is populated by helium-atmosphere white dwarfs polluted by optically undetectable traces of carbon \citep{blouin2023,camisassa2023}.

Knowing that the vast majority of DC white dwarfs contain non-trivial amounts of carbon in their atmospheres, it is logical to ask how that should affect the determination of their atmospheric parameters. Motivated by the \textit{Gaia} bifurcation, it has become common place since the work of \cite{bergeron2019} to include a trace of hydrogen when modelling helium-atmosphere DC white dwarfs \citep{kilic2020,mccleery2020,tremblay2020,caron2023,obrien2023}. When determining the parameters of a DC star using optical (and optionally, infrared) photometry, we believe that this approach remains justified. If we exclude the UV region, the first-order effect of the addition of traces of carbon or hydrogen in helium atmospheres is in both cases an increase of the He$^-$ free--free opacity. Carbon opacities also play a role (see Figure 3 of \citealt{blouin2023}), but it seems pointless to worry about this effect, as without UV data the exact carbon abundance in any given DC white dwarf is unknown. However, if UV observations are used in the fitting process, then it becomes crucial to include carbon in the models as absorption from carbon lines significantly alters the UV spectral energy distribution.

UV spectroscopy of DC white dwarfs would provide definitive evidence for the conclusion of this work by resolving carbon spectral lines. Among all confirmed DC white dwarfs in the 8000--11{,}000\,K temperature range, we found usable UV spectra in the \textit{International Ultraviolet Explorer} (IUE) and \textit{Hubble Space Telescope} archives for only two objects. Remarkably, carbon features are detected in both cases.\footnote{Both objects were already identified as carbon polluted in previous studies (GD 1072, \citealt{fey1996}; GD 84, \citealt{weidemann1995}). Note that LAWD~71 also has an IUE spectrum with carbon features and is classified as a DC white dwarf in the literature \citep{wegner1983}. However, inspection of its SPY survey spectrum \citep{napiwotzki2020} revealed the presence of Swan bands, and we therefore discarded it from our sample.} While this sample size is very small, it should motivate new UV spectroscopic observations of DC white dwarfs.

\section*{acknowledgements}
SB is a Banting Postdoctoral Fellow and a CITA National Fellow, supported by the Natural Sciences and Engineering Research Council of Canada (NSERC). AB is an NSERC Postdoctoral Fellow. This work is supported in part by NASA under grant 80NSSC22K0479, and by NSF under grants AST-1906379 and AST-2205736. This work has made use of the Montreal White Dwarf Database \citep{dufour2017}. This work has benefited from discussions at the KITP program ``White Dwarfs as Probes of the Evolution of Planets, Stars, the Milky Way and the Expanding Universe'' and was supported in part by the National Science Foundation under Grant No. NSF PHY-1748958. This research received funding from the European Research Council under the European Union’s Horizon 2020 research and innovation programme number 101002408 (MOS100PC).

This work contains results from the European Space Agency (ESA) space mission \textit{Gaia} (\url{https://www.cosmos.esa.int/gaia}). \textit{Gaia} data are being processed by the \textit{Gaia} Data Processing and Analysis Consortium (DPAC). 

\section*{Data availability}
All observational data used in this work is publicly available on the {\it Gaia} archive, GALEX archive and the Montreal White Dwarf Database.

\bibliographystyle{mnras}
\bibliography{references}

\bsp
\label{lastpage}

\end{document}